\documentclass{aa}
\usepackage{graphicx}
\usepackage{txfonts}
\usepackage{natbib}
\bibpunct{(}{)}{;}{a}{}{,}
\newcommand{\msol}{\mbox{M}_{\odot}}

\newcommand{\teff}{T_\mathrm{eff}}
\newcommand{\logg}{\log\left(g\right) }
\newcommand{\logy}{\log\left(y\right) }

\newcommand{\hal}{\mbox{H}\alpha}

\newcommand{\Ha}{H$\alpha$}
\newcommand{\tefflogg}{$\teff$-$\logg$}
\newcommand{\tefflogy}{$\teff$-$\logy$}

\newcommand{\hei}{He~{\sc i}}
\newcommand{\heii}{He~{\sc ii}}
\newcommand{\ciii}{C~{\sc iii}}
\newcommand{\civ}{C~{\sc iv}}
\newcommand{\niii}{N~{\sc iii}}

\begin{document}
\title{Hot subdwarfs from the ESO Supernova Ia Progenitor Survey\thanks{Based
  on observations collected at the Paranal Observatory of the European
  Southern Observatory for program No. 165.H-0588(A) and 167.D-0407(A).} 
}

  \subtitle{II. Atmospheric parameters of subdwarf O stars}

 \titlerunning{Hot subdwarfs from SPY -- II. Atmospheric parameters of sdO stars}

  \author{A. Stroeer\inst{1,2} \and U. Heber\inst{1} \and T. Lisker\inst{1,3}
  \and R. Napiwotzki\inst{1,4} \and S. Dreizler\inst{5} \and 
  N. Christlieb\inst{6} \and D. Reimers\inst{6}}

\offprints{U.~Heber\\
\email{heber@sternwarte.uni-erlangen.de}}

\institute{
  Dr. Remeis$-$Sternwarte Bamberg, Astronomical Institute of
  the University of Erlangen$-$N\"urnberg, Sternwartstra\ss e 7,
  D$-$96049 Bamberg, Germany \and
  School of Physics and Astronomy, University of Birmingham, Edgbaston,
  Birmingham B29 4PT, UK \and
  Astronomical Institute, Dept.\ of Physics and Astronomy,
  University of Basel, Venusstrasse 7, CH$-$4102 Binningen, Switzerland \and
  Centre for Astrophysics Research, University of Hertfortshire, College
  Lane, Hatfield AL10 9AB, UK \and
  Institut f\"ur Astrophysik,
  University of G\"ottingen, Friedrich-Hund-Platz 1, D$-$37077 G\"ottingen,
  Germany \and
  Hamburger Sternwarte, Universit\"at Hamburg, Gojensbergweg 112, 21029 Hamburg, Germany
}

\date{Received date / Accepted date}

\abstract{}{
We address the origin and evolutionary status of hot
subdwarf stars by studying the optical spectral properties of 58 subdwarf O
(sdO) stars. Combining them with the results of our previously studied 
subdwarf B (sdB) stars, we aim at investigating possible evolutionary links.
}
{We analyze high-resolution 
(R$>$18000), high-quality optical spectra of sdO stars obtained with the ESO 
VLT UVES echelle spectrograph in the course of the ESO Supernova Ia Progenitor 
Survey (SPY). Effective temperatures, surface gravities, and photospheric 
helium abundances are determined simultaneously by fitting the profiles of 
hydrogen and helium lines using dedicated synthetic spectra calculated from an 
extensive grid of NLTE model atmospheres.
}
{
We find spectroscopic or photometric evidence for cool companions to 
eight sdO stars as well as a binary consisting of two sdO stars. 
A clear correlation between helium abundances and the presence of
carbon and/or nitrogen lines is found: below solar helium abundance, no sdO star
shows C or N lines. In contrast, C and/or N lines are present in the
spectra of all sdO stars with supersolar helium abundance. We
thus use the solar helium abundance to divide our sample into
\emph{helium-deficient} and \emph{helium-enriched} sdO stars.
While \emph{helium-deficient} sdO stars are scattered in a wide range of the 
\tefflogg-diagram, most of the 
\emph{helium-enriched} sdO stars cluster in a narrow region  at
temperatures between 40\,000 and 50\,000K and gravities between $\log$ g=5.5 
and 6.0.
}
{
An evolutionary link between sdB stars and sdO stars appears plausible only for
the \emph{helium-deficient} sdO stars. They probably have evolved away from the extreme
horizontal branch, i.e. they are the likely successors to sdB stars. 
In contrast, the atmospheric properties
of \emph{helium-enriched} sdO stars cannot be explained with canonical
single star evolutionary models. Alternative scenarios for
both single star (late hot flasher) as well as for binary evolution 
(white dwarf merger; post-RGB evolution) fail to 
reproduce the observed properties of \emph{helium-enriched} sdO stars in detail. 
While we regard the post-RGB scenario as inappropriate, the white dwarf merger 
and the late hot flasher scenarios remain viable to explain the origin of
\emph{helium-enriched} sdO stars.
} 


\maketitle



\section{Introduction \label{sec:int}}

Hot subluminous stars are an important population of faint blue stars at high
galactic latitudes closely related to the horizontal branch.
A proper spectral classification of hot subluminous stars is rendered
difficult by the diversity of the helium line spectra. They can roughly be
grouped into 
the cooler sdB stars, whose spectra typically display no or only weak helium
lines, and the hotter sdO stars, which have a higher helium abundance on
average and can even be dominated by helium.
The former have recently been studied extensively because they are
sufficiently common to account for the UV excess observed in early-type
galaxies. Pulsating sdB stars are important tools for asteroseismology
\citep{char04} and sdB stars in close binaries may qualify as Supernova Ia
progenitors \citep{maxt00,geie06}. 

Subluminous B stars have been identified as Extreme Horizontal Branch (EHB)
stars
\citep{heb86}, i.e. they are core helium burning stars with hydrogen envelopes 
too thin to sustain hydrogen burning (unlike normal HB stars).
Therefore they evolve directly to the white dwarf cooling sequence avoiding 
the Asymptotic Giant Branch (AGB).
While the sdB stars spectroscopically form a homogenous class, a large variety 
of spectra is observed among sdO stars \citep{heb92,heb06}. 
Most subluminous B stars are helium poor, whereas only a relatively
small
fraction of sdO stars are.

Ever since the pioneering work by \citet{green74},
the helium-rich sdO stars were believed to be linked to
the evolution of the hydrogen-rich subluminous B stars. 
Any evolutionary link between subluminous B and O stars, however, is
difficult to explain since the physical processes
driving a transformation of a hydrogen-rich star into a helium-rich one
remain obscure. The convective transformation has been explored by 
\citet{wese81} as well as by \citet{grot85}. While the former
found helium convection to occur even at subsolar helium abundances which
mixes helium from deeper layers into the photosphere, the
latter concluded that a helium driven convection zone develops only in
helium-rich atmospheres. If the latter is true, convective transformation
would not work.
Non-standard evolutionary models were introduced to explain the formation of
sdO stars \citep[e.g.][]{swei97,bro01,moe04}. In particular, 
the {\it late hot flasher scenario} predicts that the core helium flash may
occur
when the
star has already left the red giant branch (RGB) and is approaching the white 
dwarf cooling sequence (delayed He core flash). During the flash, He 
and C are dredged-up to the surface. Hydrogen is mixed into deeper layers and 
burnt. The
remnant is found to lie close to the helium main sequence, i.e. at the very end
of the theoretical extreme horizontal branch.

The fraction of sdB stars in short period binaries (periods less than ten days) is high.
\citet{max01} found 2/3 of their sdB sample (drawn from the Palomar Green
survey) to be such binaries, whereas 
 somewhat lower fractions of 40\% and 48\% were found recently for 
  the SPY sample \citep{napi04} and for a sample drawn from the Edinburgh-Cape
  catalog \citep{mora06}, respectively.
Obviously, binary evolution plays an important role for the formation of sdB 
stars and possibly also for that of the sdO stars. 
A recent population synthesis study \citep{han03} 
identified three channels 
to form sdB stars:
(i) one or two phases of common envelope evolution,
(ii) stable Roche lobe overflow, and
(iii) the merger of two helium-core white dwarfs.
The latter could explain the population of single stars.
The simulations by \citet{han03}  cover the observed parameter range of sdB 
stars \citep[see][henceforth paper~I]{lis05}.

The ESO Supernova Ia Progenitor SurveY \citep[SPY,][]{spy0} has obtained
VLT/UVES spectra for over 1000 white dwarf candidates to test
possible scenarios for type Ia supernovae by searching for double
degenerate white dwarf binary systems close to the Chandrasekhar mass
limit. Many of the target stars of SPY came from the Hamburg ESO survey 
\citep{wiso96}. SPY also observed 137 hot subluminous stars that entered the
target sample because they were previously classified as white dwarfs. 
76 of these stars were now classified as sdB/sdOB, and 58
as O-type subdwarfs \citep{chri01,lis03}.  

The data for our hot subdwarf sample are of unprecedented quality and
homogeneity. Spectral analyses of all sdB stars from that sample have
already been presented in paper~I in order to test evolutionary
models, in particular the binary population models of \citet{han03}. 
Two diagnostic tools -- the effective temperature
($\teff$) vs.\ surface gravity ($\logg$) diagram and the cumulative
luminosity function -- yielded 
conflicting results. Moreover, the models of \citet{han03} predicted some EHB
stars to be hotter than the sdB stars contained in the sample of
paper\,I. This led us in paper I to the conclusion that their sample of
hot subdwarfs may not be sufficiently complete to describe the whole
parameter range covered by the simulations, and that it needs to be
complemented by a similar analysis of subdwarf O stars.

In this paper we focus on 
the subdwarf O spectra from
the SPY sample and present the spectral analysis of the high
resolution spectra using
state-of-the-art NLTE model atmospheres.
This enables us to address the
still open question about the evolutionary status of hot subdwarfs.
We combine our results for the SPY sdO stars with those for the SPY
sdB stars from paper\,I to cover the entire stellar atmospheric
parameter range with the same high accuracy. This allows for the
first time a detailed comparison with evolutionary model predictions,
particularly in the regime of hotter temperatures and higher helium
abundances than in previous studies of sdB stars alone. 

The paper is organised as follows. In section~\ref{sec:data} we
outline the available data and explain our criteria for spectral
classification. Section~\ref{sec:nlte} briefly describes
the construction of a new NLTE model atmosphere grid.  
In section~\ref{sec:results} we detail the derivation
of stellar parameters by fitting model atmospheres to observed spectral
lines and present the results in section~\ref{sec:result2}. 
Various evolutionary scenarios are tested in
section~\ref{sec:obstheo}, leading to a discussion and summary in 
section~\ref{sec:discuss}. 

\section{Observation, data reduction and spectral classification
\label{sec:data}}

\subsection{Observation and data reduction} 

Observations were obtained at the ESO Very Large Telescope with UT2
(Kueyen) equipped with the UV-Visual Echelle Spectrograph (UVES). A
slit width of $2\farcs1$ was used, resulting in a resolving power
of $18\,500$  (spectral resolution of 0.36\,\AA\ at $\hal$) or better. 
Wavelength coverage
of  3300-6650\,\AA\ is achieved, with gaps at 4500-4600\,\AA\ and
5600-5700\,\AA\ \citep[][]{spy0,koe01}. For most of the stars, two
exposures in different nights were taken, since SPY was originally
intended to search for radial velocity (RV) variable objects. The
spectra were then reduced with a procedure developed by C.~Karl using
the ESO MIDAS software package, partly based on the UVES pipeline developed at
ESO. 

\subsection{Spectral classification}\label{sec:class}

For many years it has been attempted to establish a consistent classification
of hot subdwarfs 
\citep{green86,jef97,dril03} that applies to lower resolution spectra.
A transfer of these schemes 
to spectra of higher resolution would be desirable. However,  
in case of our UVES spectra, we lack the \heii\ 4542\AA\ line due to a 
wavelength gap between the blue
and the red arm of the spectrograph.
Since this line
plays a crucial role in the 
classification scheme of \citet{dril03}, we cannot apply this scheme to our
data. Instead we aim for a less detailed
classification strategy.

In the process of classifying all SPY spectra \citep{lis03}, 
we therefore just separated the spectra 
into hot white dwarfs (of various subtypes), sdB, He-sdB, sdO, and He-sdO stars by visual 
inspection and 
comparison with spectra of prototypical stars. 
The terms
He-sdB/He-sdO were introduced to mark extremely helium-rich sdB/sdO
stars \citep[]{ahm03,ahm05}. 
Note that we did not use 
the intermediate class of sdOB stars introduced by \citet{bas75} -- we 
subsumed them in the term sdB. 
For our programme stars in the present paper we continue to use only the 
general term sdO , in order
to prevent any bias from qualitative visual classification entering 
our quantitative study of atmospheric parameters.

It is worthwhile to note that our classification and the one
of \citet{green86} do not agree as becomes apparent from 
several stars that we have in common 
(see Table~\ref{tab_newgrid} of this paper and Table~1 of paper~I). 
Seven programme 
stars (four sdB and three sdO) have been misclassified as white dwarfs by 
\citet{green86}. While none of the sdB stars in paper~I has been classified
as sdO star in the PG survey, six of our sdO stars are classified as 
sdB in the PG catalog. Five of them turn out to be helium deficient (see section
\ref{sec:result2}). It is
likely that the helium deficient sdO stars are subsumed in the sdB class of the
PG catalog. 
Hence the fraction of sdO stars is underestimated by \citet{green86} .

The high spectral resolution of the UVES spectra does not only allow to
identify and measure hydrogen and helium lines, but also the metal line
spectrum can be investigated. Lines from highly ionized carbon and 
nitrogen
are found in normal O-type stars.
For a refined spectral classification we thus used the presence or absence of
characteristic absorption lines of carbon and nitrogen
to divide our sample
into several subclasses ('CN-scheme'). 
 For carbon we used the strongest lines of
\ciii\ (4650\AA, multiplet\,1,
4070\AA, mult.\,16, and 4186.9\AA\ of mult.\,18) and of \civ\ 
(5801.33\AA\ \&  5811.98\AA,
mult.\,1, and 4658.3\AA, mult.\,8). For
nitrogen the strongest lines of \niii\ are used 
(4640\AA, mult.\,2  and 4858\AA, mult.\,9, 4379.11\AA, mult.\,18,)  

We adopt the following classification scheme:
\begin{itemize}
\item
Stars with \ciii\ and/or \civ\ visible in their spectra but with no \niii\ are
classified as 'C'.
\item
Stars with \niii\ and \ciii\ and/or \civ\ in their spectra were classified as 'CN'.
\item
The absence of \ciii\ and \civ\ and the presence of \niii\ leads to the
classification 'N'.
\item
With no lines of \ciii, \civ, and \niii\ visible a star was classified as '0'.
\end{itemize}
This is not an optimized, quantitative classification scheme, but
it is intended to demonstrate that even such a qualitative subdivision already 
shows a correlation with the atmospheric parameters of our stars, as we
shall outline below.
The resulting classifications are presented in
Table~\ref{tab_newgrid}.
We checked whether the presence or absence of CN might be significantly 
influenced by the signal-to-noise ratio (S/N), 
and found that the S/N distribution of our spectra does 
not differ for different CN-classes 
at low and intermediate S/N values, for which it could be crucial. Hence the 
CN classification scheme is not significantly affected by noise.

\subsection{Search for companions of sdO stars}

 In paper~I we found a significant fraction of sdB stars with cool companions
 (24 out of 76 sdB stars).
 Therefore we searched for spectroscopic and photometric evidence for such
 companions amongst our sdO stars.
 
 Only for one subluminous O star, {\bf HE~1502$-$1019} (alias PG~1502$-$103; 
 EC~15026$-$1019), 
 spectroscopic signatures for a cool companion were found, like the the G-band,
 and the Mg-triplet. \citet{fer84} classified the companion star as spectral
 type K0.5. The weakness of the helium lines implies a very low 
 helium abundance of about He/H=0.001. We estimate the effective temperature 
 from a
 comparison to synthetic spectra (see section~\ref{sec:nlte}) 
 to be near 45\,000K . 
 
 {\bf HE~0301$-$3039} turns out to be radial velocity variable and its
 spectrum also is composite; it indicates
 that HE~0301$-$3039 is a binary    
 consisting of two sdO stars with spectra dominated by 
 helium lines, the first such system found \citep{lis04}.
 
 {\bf HE~1200$-$1924} (alias EC~12001$-$1924; Feige~54) shows a 
 helium dominated
 absorption line spectrum typical of a sdO star.
However, Balmer emission lines are superimposed -- 
strong and broad H$\alpha$, strong but narrow H$\beta$ 
and weak H$\gamma$ and H$\delta$ emission. These emissions do not originate from
the photosphere of the sdO star but may stem from a companion star.
No spectral absorptions typical for a cool companion can be found. The $B-J$
and $J-K$ colours, however, are unusual for a sdO star (see below).    
 
 For the above objects too few spectra are at hand to disentangle the 
 individual spectra and the
 quantitative spectral analysis of these objects has to be postponed until they
 become available.
 
 We also searched for photometric evidence for companions by complementing the
  B-band fluxes with IR measurements from the 2MASS \citep{2MASS} 
  and DENIS\footnote{http://cdsweb.u-strasbg.fr/denis.html} catalogs.
Photoelectric B magnitudes (accurate to better than $\pm 0\fm05$) 
were taken from 
the subdwarf data base 
\citep{oest06}. If the former were unavailable, 
B magnitudes are from the Hamburg-ESO survey 
\citep[][accurate to $\pm 0\fm2$]{wiso00}.

\begin{figure}
\resizebox{\hsize}{!}{\includegraphics{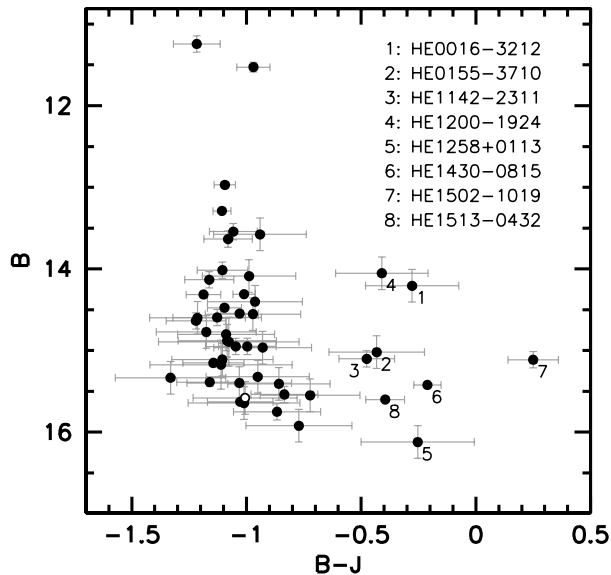}}
\caption{
Colour-magnitude diagram, showing B magnitude versus
optical-to-near-infrared colour (B-J). 
Numbers denote objects that lie off the bulk of 'normal' sdOs,
implying a cool companion. The sdO-sdO binary HE0301-3039 is shown as open
circle. HE~1502$-$1019 and HE~1200$-$1924 are
composite spectrum objects (see text).} 
\label{fig:bmj}
\end{figure}

J magnitudes and corresponding errors are provided by
 2MASS  for 51 stars, and by 
 DENIS for one
star.  
Only the brightest stars were detected in the K band.  
A color-magnitude diagram is displayed in Fig.~\ref{fig:bmj}. Interstellar
extinction and reddening have been corrected for using the maps of 
\citet{schl98}. As most of the stars are located at high galactic latitudes,
$E(\rm {B-V})$ is small ($\le 0\fm12$) except for the low latitude star  
HZ~1. For the latter, the reddening as inferred from the Schlegel maps 
is overestimated, since the star lies near the Galactic plane (z$\approx$80pc). 
Instead, we compare the measured $B-V$ to model predictions,
yielding $E(\rm {B-V})=0\fm25$.

In paper I we found that sdB stars with spectroscopic signatures of cool
companions are redder than $B-J \approx -0\fm4$. Similarly, eight sdO
stars in Fig.~1 lie at $B-J > -0\fm5$ and clearly separate from the bulk
of sdOs. We number those objects in the figure. 
Among them is HE~1200$-$1924, whose emission lines provide 
additional support for the existence of a cool companion to that sdO star.

Only for 12 stars, K band magnitudes to better than $\pm 0\fm2$ are available.
\cite{stark03} studied the distributions of $J-K$ colours of sdB and sdO stars 
and find both to be double-peaked, the sdO stars at $J-K=-0\fm2$
and $+0\fm2$,
respectively, indicating that the latter sdO stars have cool companions. 
$J-K$ measurements are available for five of the eight stars highlighted in 
Fig.~\ref{fig:bmj}. All five have positive $J-K$ indicating the presence of 
cool companions.  

Judged from our experience with sdB stars, we can assume the blue 
spectra of these stars not to be affected by light from the cool companion.
Those stars are marked in Table~\ref{tab_newgrid} and the quantitative spectral
analysis is performed
in the same way as for the other stars (see section~\ref{sec:results}).

\section{Stellar atmosphere modeling}\label{sec:nlte}

High quality optical spectra like those from SPY demand accurate
theoretical counterparts as input for fitting routines.  

Only few detailed quantitative spectral analyses have been carried out for
the sdO class because of the complexity of their spectra.
In addition deviations from local 
thermodynamic equilibrium (NLTE) have to be taken into account, 
because of their high effective temperatures, whereas the LTE assumption is
reasonable for sdB stars.
Therefore, only few quantitative spectral analyses have been published
\citep[]{dre90,the94}, with some conflicting results 
becoming apparent.
The reliability of the NLTE calculation depends strongly on the quality and
sophistication of the model atom and the atomic data used.
The helium model atoms
are of utmost importance for the modeling of sdO atmospheres. We used 
more detailed model atoms for \hei\ and \heii\ than in \citet{dre90}
and constructed a new grid of atmospheric models
and synthetic spectra using a state-of-the-art NLTE model atmosphere code.

An extensive grid of NLTE atmosphere models was calculated using the
latest version of the PRO2 code \citep{wern99} that employs a
new temperature correction technique \citep{drei03}. A new detailed
model atom for helium appropriate for the sdO temperature regime was
constructed.
2700 partially line blanketed NLTE
model atmospheres consisting of hydrogen and helium
were calculated
resulting in
a grid of unprecedented coverage and resolution, extending from
30\,000~K to
100\,000~K in T$_\mathrm{eff}$. The gravity ranges from $\log{g}$=4.8 to 6.4 
and the helium abundance
from $\log{N_{He}/N_H}$=$-$4 to $+$3 in order to
match the diversity of observed spectra. 
The step sizes are 2\,000~K from
30\,000~K to 52\,000~K and 5\,000~K from 55\,000~K to
100\,000~K; 0.2 and $\sim$0.5 dex, respectively.

As a test we carried out the spectral analyses of several programme stars
using both the old grid of  \citet{dre90} and the new one. The synthetic spectra from
the new models match the observed profiles much better in all cases.

The differences of the current model grid compared to the one used by
\cite{dre90} can be traced back to two major improvements: The
resonance lines of \ion{He}{II} could only be treated in detailed
balance in the previous model grid. This eliminates the transition from
the statistical equilibrium equations resulting in a more stable
numerical behavior. With an improved numerical stability of the code
this approximation is no longer necessary. Since the resonance lines of
\ion{He}{II} have a strong influence on the structure of the atmosphere
of sdO stars, this improvement is the dominant effect. The second
improvement since \cite{dre90} is the treatment of the level
dissociation according to \cite{humi88}. 
This mainly affects
higher line series members and is therefore important for a  precise
gravity determination.


\section{Spectral analysis: techniques}\label{sec:results}

\subsection{Line profile fitting \label{subsec:fit}}

Atmospheric parameters ($\teff$, $\logg$,
$\logy$, y=He/H by number) were 
determined simultaneously by fitting the synthetic
spectra  to observed hydrogen and 
  helium line spectra using a $\chi^2$-procedure 
  \citep{napfit}.

\begin{figure}
\resizebox{\hsize}{!}{\includegraphics{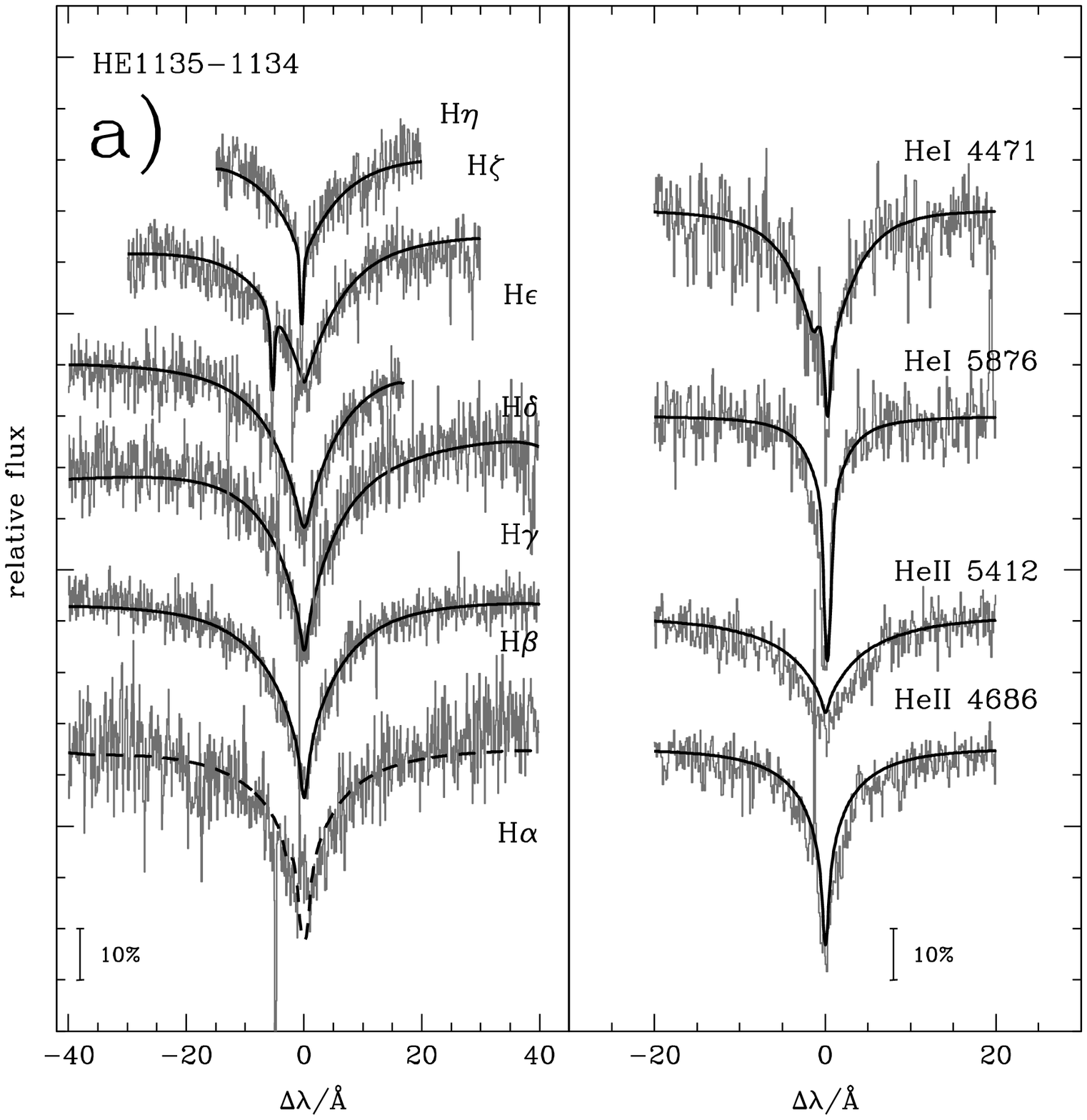}}
\resizebox{\hsize}{!}{\includegraphics{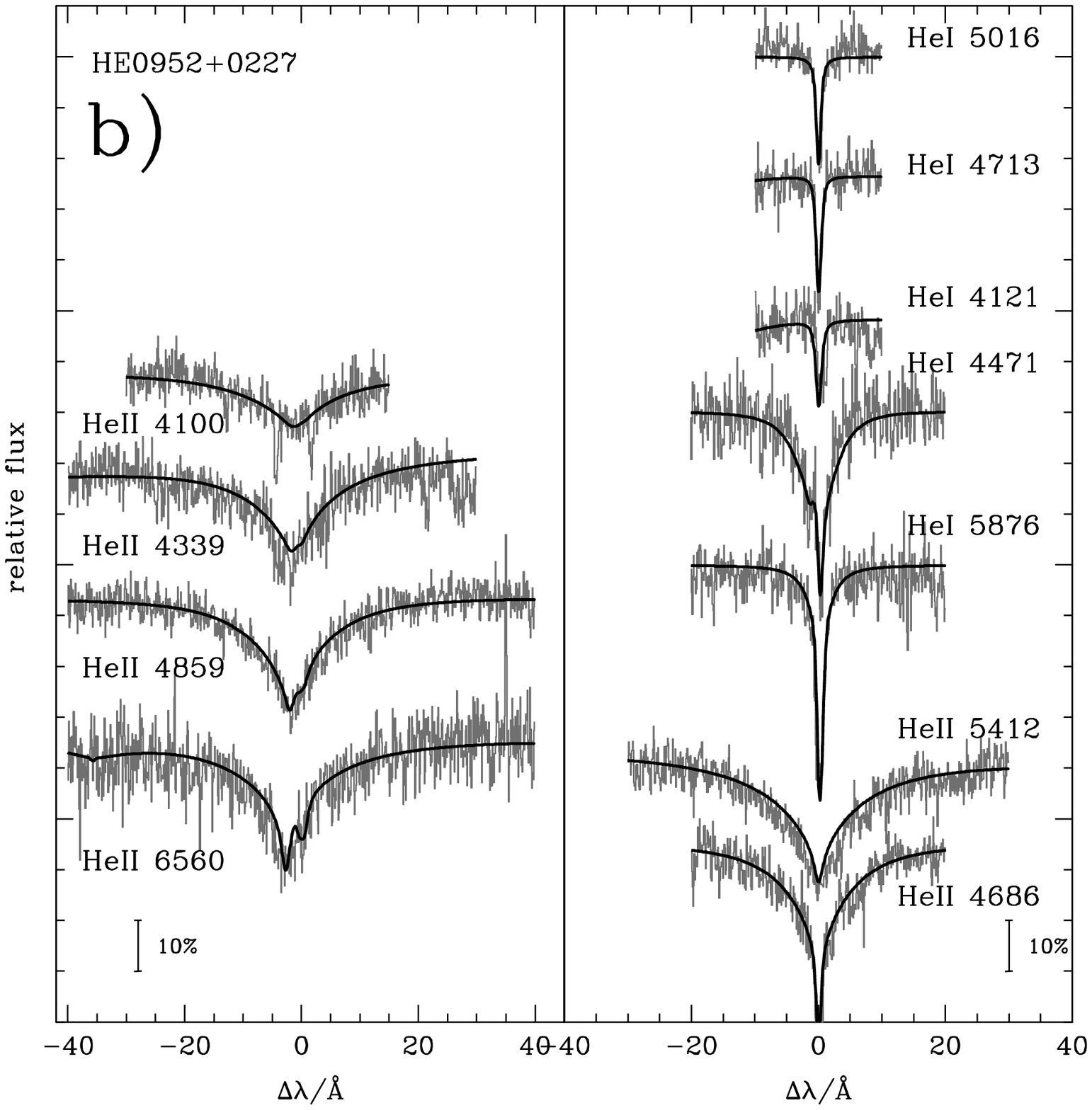}}
\caption{Sample fit of two sdO star spectra (grey) using the new grid of synthetic 
NLTE spectra as described in Section~\ref{sec:nlte}. 
In HE~1135$-$1134 (top panel) hydrogen
is more abundant than helium, whereas in HE~0952+0227 (bottom panel) 
hydrogen is a trace element only.
The model spectra yielding the best fits are shown as black lines. 
 For details see text.
}
\label{fit}
\end{figure}
For each line, the continuum level is determined and
normalized to 1, in order to compare it to the synthetic spectrum. For sdO 
stars with strong hydrogen Balmer lines, 
the \Ha\ was excluded in the parameter
determination because peculiarities in the \Ha\ line core were often reported, 
perhaps caused by
stellar winds \citep[]{hebwind}. Nevertheless \Ha\ was kept in the final
plot for visually examining any significant deviations from the model
profiles.
Those sdO stars showing no evidence for hydrogen from the visual 
inspection of the \heii\ Pickering decrement required a modified 
fitting strategy.
If hydrogen were present as a trace element it could be
best detectable as a contribution of \Ha\ to
the \heii\ Pickering line 6560\AA\ but may be too weak to be visible to the 
eye. In a first step $\teff$
and $\logg$ are derived by a fit using all  
lines including 6560\AA. Then a second fit is carried out to
derive the hydrogen abundance keeping $\teff$ and $\logg$ fixed from the
first run.
In a third step new values for $\teff$ and
$\logg$ are derived while keeping the hydrogen abundance fixed.
If the new values differ by more than 0.5\%,
we repeat the procedure until convergence is achieved.

Two stars in our sample (DeHt~2 and K2-2) have already been analysed 
by \citet{napatm} with similar NLTE model atmosphere techniques 
as presented here, and are known to be central stars of planetary nebulae
(CSPN).
DeHt~2 was classified as a high luminosity CSPN and 
$\teff$=117000~K, $\logg$=5.64 and $\logy$=$-$0.79 
 were derived, whereas
 K~2-2 was classified as a peculiar CSPN and 
 $\teff$=67000~K, $\logg$=6.09 and $\logy$=$-$1.55 resulted. These stars will not
 be discussed further in this paper.

For 46 out of the remaining 53 stars the fit procedure converged to 
convincing line profile fits (as displayed in Fig.~\ref{fit} for two typical
cases). 
As can be seen the agreement between the best fit synthetic spectra
and the observed ones is excellent.

However, for seven stars the fit procedure failed due to various
reasons.
We shall discuss those stars in some detail below.

{\bf HE~1349$-$2320} (alias EC~13494$-$2320) and {\bf HE~1355$-$0622} 
(PG~1355$-$064): weak He~I lines and strong 
\heii\ lines indicate high temperatures. However, Balmer and \heii\ lines cannot 
be matched simultaneously. The H$\alpha$ lines display weak emission cores 
not reproduced by our models.

{\bf HE~1512$-$0331} \citep[see][]{chri01} displays very broad Balmer and 
\heii\ lines with emission
cores of \heii, 6560\AA\ and 4686\AA. 
Its gravity is too high for our model grid. Extrapolation suggests that 
its temperature may be near 80000~K and $\logg$ slightly below 7. We regard
the star as a DAO white dwarf. 

{\bf HE~1518$-$0948} \citep[alias PG~1518$-$098, see][]{moe90}: 
Weak He~I lines indicate high temperature, while Balmer
and \heii\ lines are narrow but strong indicating that it is a hot, helium-rich
sdO star of low gravity. In fact, the attempts to fit its spectrum showed 
that the gravity must be lower than the lower limit of the grid, i.e. 
$\logg$=4.8 at a $\teff$ near 60000~K.

{\bf HE~2305$-$1155} 
displays strong and broad He~I lines 
indicating a low $\teff$. Trials to match its spectrum indicated a very 
high gravity beyond the limit of our model grid 
(i.e. $\logg>$6.4). 

{\bf HE~1008$-$179} \citep[see][]{chri01} is a very hot, high gravity sdO star showing a H$\alpha$ 
emission core. Individual Balmer lines can not be matched simultaneously
\citep[the so-called Balmer line problem, see][]{napatm}.
Attempts to fit its spectrum indicated that its parameters are probably beyond
the high temperature, high gravity limit of the model grid, i.e. exceeding 
$\teff$=100000~K and $\logg$=6.4, and the star should be classified as an 
extremely hot 
white dwarf.
 
{\bf EC~11481$-$2303} has already been been analysed by \citet{stys00} who 
derived $\teff$=41790~K, $\logg$=5.84, $\logy$=$-$1.85. The authors were  
unable to match the UV spectral energy distribution. 
The Balmer line problem is obvious in our optical spectra from the fit, 
which indicated a 
higher temperature ($\approx$ 50000~K), lower helium content ($\approx -$2.8)
but similar gravity than found by \citet{stys00}.
The Balmer
line problem in very hot white dwarfs was traced back to metal line blanketing 
by \citet{wern96}. 
\citet{hamm03} analysed the UV spectrum of EC~11481$-$2303 and derived 
very large iron and nickel abundances of 10 to 100 times solar, 
causing a strong line blanketing effect. Hence, the
neglect of metal line blanketing is the likely cause of our failure to 
match the optical spectra of this star.

Because of the peculiarities described and problems encountered in the fit
procedure, we do not include these stars in the further discussion and are left
with a working sample of 46 stars.

Since the SPY project was originally intended to search for radial
velocity variable stars,
two or more exposures of each star, if available. Thus the final values were
calculated as the average of the individual fit results, weighted
with the $S/N$ of the corresponding spectra. The final
values are presented in 
Table~\ref{tab_newgrid}. 
We note, that for seven stars of the
working sample
only a single (useful) exposure was available and denote them by 'S'
in Table~\ref{tab_newgrid}. 
Effective temperatures 
range from
36\,000\,K to 78\,000\,K, gravities from $\logg$=4.9 to 6.4, and 
helium-to-hydrogen ratios from 0.0003 to more than 1000.

By adopting a canonical mass of 0.47~$\msol$ for our programme stars, we
can further derive the luminosity in solar units (Table
\ref{tab_newgrid}). 

\begin{table*}
\caption{Results of the quantitative spectral analysis of the ESO SPY sdO 
stars 
sorted by increasing helium abundance ($\logy$). 
    The results are
mean values from two or more exposures,
if available. ''Outliers'', i.e. values with relative errors between 
two or more exposures above the 2-$\sigma$ level of the global relative error, 
are flagged by an asterisk (see section \ref{subsec:errors}). 
The CN-classification is given as described in
    Section~\ref{sec:class}. B magnitudes are from the Hamburg ESO
    objective
    prism survey  \citep[photographic, accurate to $\pm 0.2$\,mag, 
    see][]{wiso00} if not specified otherwise.}

\label{tab_newgrid}

    \centering
         \begin{tabular}{llllllllll}
            \hline
            \hline
            \noalign{\smallskip}
            Object&ICRS 2000.0&B&M$_V$\hspace*{1cm}&$\teff$ & $\logg$ & $\logy$ & $\log(L)$ &CN & Rem.\\
             & &mag&mag &[K] & $[\mathrm{cm\ s^{-2}}]$ &  &
$[\mathrm{L_{\odot}}]$ &&   \\
            \noalign{\smallskip}
            \hline
            \noalign{\smallskip}
	    \multicolumn{10}{l}{helium deficient:}\\
HE~1059$-$2735 &11:01:24.95 $-$27:51:42.9&15.21$^{\rm a}$ &3.11& 40966 & 5.38 
& $-$3.62 & 2.16 &0& 2\\
HE~1130$-$0620 &11:32:41.60 $-$06:36:54.4&15.76$^{\rm a}$ &4.05& 48122 & 5.84 
& $-$3.07 & 1.98 &0& 1\\
HE~1237$-$1408 &12:39:56.64 $-$14:24:48.4&15.97$^{\rm a}$ &3.02& 39683 &5.33 & 
$-$2.99 & 2.15 &0& 1,2,5,8\\
HE~1318$-$2111 &13:21:15.66 $-$21:27:18.5&14.48$^{\rm a}$ &3.31 & 36254$^{\ast}$& 5.42 
 & $-$2.91 & 1.92 &0& 2,5\\
HE~1115$-$0631 &11:18:11.69 $-$06:47:33.2&14.77$^{\rm a}$ &4.17 & 40443 & 5.80 
& $-$2.59 & 1.72 &0& 1\\
PG~0026$+$136  &00:28:52.33 +13:54:45.8&15.74$^{\rm 1}$ &3.17 & 38830 & 5.38 & 
$-$2.39 & 2.07  &0&  1\\
EC~09445$-$0905  &09:47:03.39 $-$09:19:50.5&15.69$^{\rm a}$ &4.59 & 73862 & 6.22
& $-$2.08 & 2.28  &0& 2,4  \\
HE~1423$-$0119 &14:25:51.29 $-$01:33:17.4&16.56$^{\rm a}$ &3.12 & 52662 & 5.50 
& $-$1.61 & 2.48 &0& 1\\
HE~1513$-$0432 &15:16:19.17 $-$04:43:58.0&16.04$^{\rm a}$ &4.41 & 42699 &5.92 &
 $-$1.36$^{\ast}$ & 1.69 &0& 1,7\\
HE~0040$-$4838 &00:42:31.08 $-$48:22:16.2&16.07&3.79 & 41823 & 5.67 
& $-$1.35 & 1.91 &0& \\
HE~1047$-$0637 &10:50:28.79 $-$06:53:25.9&14.42$^{\rm a}$ &1.49 & 60650 & 5.03 
& $-$1.34 & 3.39 &0& 1,5\\
HE~1356$-$1613 &13:59:12.52 $-$16:28:01.8&16.18&4.11 & 55925$^{\ast}$ & 5.93 
& $-$1.30 & 2.15 &0& 5 \\
HE~1106$-$0942 &11:09:08.22 $-$09:58:48.6&16.34&4.92 & 79742$^{\ast}$ & 6.40 
& $-$1.03 & 2.27 &0&1\\
            \hline
	    \multicolumn{10}{l}{helium enriched:}\\
HE~1238$-$1745 &12:41:01.16 $-$18:01:59.0&14.31$^{\rm a}$ &3.83 & 38219 & 5.64
 & $-$0.55 & 1.78 &N& 2,5\\
HE~1258$+$0113 &13:00:59.23 +00:57:11.7&16.23$^{\rm b}$ &3.80 & 39359 & 5.64 
& $-$0.53 & 1.83 &N& 1,5,7\\
HE~2218$-$2026 &22:21:13.02 $-$20:11:17.6&16.28&4.40 & 38330 & 5.87 
& $-$0.35 & 1.56 &CN& \\
HE~1135$-$1134 &11:38:10.66 $-$11:51:03.8&15.45&3.88 & 40079 & 5.68 
& $-$0.26 & 1.82 &N& 1\\
HE~1136$-$2504 &11:39:10.21 $-$25:20:55.5&13.83$^{\rm a}$ &4.25 & 41381 & 5.84 
& $-$0.16 & 1.72 &N& 2,5\\
HE~1310$-$2733 &13:12:50.65 $-$27:49:03.1&14.38&3.76 & 40000 & 5.63 
& $-$0.15 & 1.87 &N& 2,5\\
HE~2359$-$2844 &00:01:38.48 $-$28:27:42.8&15.7 &3.85 & 38325 & 5.65 
& $-$0.15 & 1.77 &CN& 3 \\
PG~2204$+$070  &22:07:16.20 +07:18:36.0&15.74$^{\rm 4}$ &3.72 & 40606 & 5.62 
&0.07 & 1.90 &N& 1,4,8\\
HE~1256$-$2738 &12:59:01.48 $-$27:54:19.3&16.29$^{\rm a}$ &4.04 & 40029 &5.68 
& 0.09 & 1.82    & CN & 2,5 \\
HE~2203$-$2210 &22:06:29.38 $-$21:56:00.0&15.04&4.95 & 47343 & 6.20 
& 0.45 & 1.59 &CN& \\
HE~1142$-$2311 &11:44:50.15 $-$23:28:18.0&15.37$^{\rm a}$ &3.86 & 54718 & 5.80
 & 0.68 & 2.21 &C$^d$ & 2,5,6\\
HE~0111$-$1526 &01:13:38.24 $-$15:11:02.6&14.59$^{\rm a}$ &3.86 & 39152 & 6.31
 & 0.83 & 1.16 & CN & 3,8\\
HE~1251$+$0159 &12:54:08.35 +01:43:24.0&15.24$^{\rm a}$ &4.47 & 48208 & 5.98 
& 1.03 & 1.80 &C& 1 \\
HE~1511$-$1103 &15:14:17.04 $-$11:14:13.6&14.78$^{\rm a}$ &3.88 & 42298 & 5.68 
& 1.10 & 1.91 &CN& 1,5\\
HE~1430$-$0815 &14:33:36.93 $-$08:28:24.8&15.69$^{\rm a}$ &2.32 & 61011 & 5.26
 & 1.17 & 2.97 &C&1,2,7,8\\
HE~1203$-$1048 &12:05:56.59 $-$11:05:29.4&15.69&3.98 & 45439 & 5.91$^{\ast}$ 
& 1.36$^{\ast}$ & 1.91 &C& 1\\
HE~1446$-$1058 &14:49:24.49 $-$11:11:19.0&15.00$^{\rm a}$ &3.91 & 45000 & 5.76 
& 1.37 & 1.94 &CN & 2\\
HE~0342$-$1702 &03:44:58.82 $-$16:52:42.2&14.75&4.18 & 41914 & 5.78 
& 1.40 & 1.79 &N& \\
HE~0952$+$0227 &09:55:34.57 +02:12:47.9&14.72&3.84 & 46524$^{\ast}$ & 5.75 
& 1.41 & 2.00 &C$^d$ & 1,5\\
HE~2347$-$4130 &23:50:19.70 $-$41:14:01.1&15.16&4.07 & 44875 & 5.83 
& 1.44 & 1.88 &C& \\
HE~0414$-$5429 &04:15:30.23 $-$54:21:58.7&14.60&3.86 & 44678 & 5.76 
& 1.57$^{\ast}$ & 1.96 &C$^d$ & \\
HE~0914$-$0314 &09:17:15.62 $-$03:53:57.3&14.93&3.97 & 45136 & 5.79 
& 1.65 & 1.94 & C$^d$ &1\\
BPS~CS~22955$-$0024 &20:23:50.26 $-$25:08:28.8&15.50$^{\rm c}$ &4.04 & 44622 & 5.80 
& 1.72 & 1.89 &CN& 4\\
HE~0155$-$3710 &01:58:01.44 $-$36:56:21.9&15.08&4.24 & 41405 & 5.77 
&1.76 & 1.76 &N &3,7\\
HE~1136$-$1641 &11:38:54.62 $-$16:58:13.4&14.82$^{\rm a}$ &4.03 & 44646 & 5.80 
& 1.81 & 1.88 &C& 2\\
HE~0958$-$1151 &10:00:42.64 $-$12:06:00.0&13.81$^{\rm a}$ &3.29 & 44125 & 5.51$^{\ast}$ 
& 1.85 & 2.18 &C$^d$ & 1,2,4,5\\
HE~0016$-$3212 &00:18:53.22 $-$31:56:01.7&14.27&3.73 & 41674 & 5.70$^{\ast}$ 
& 2.14 & 1.97 & CN & 3,6\\
HE~0031$-$5607 &00:34:07.75 $-$55:51:05.9&15.45&5.37 & 41423 & 6.25 
& 2.26$^{\ast}$ & 1.31 &N& 8\\
HE~1316$-$1834 &13:19:16.93 $-$18:49:52.0&16.24&3.55 & 42811 & 5.56 
& 2.32 & 2.05 &N& 2,8\\
PG~2258$+$155  &23:00:57.75 +15:48:39.8&15.14$^{\rm c}$ &4.77 & 42084 & 6.08$^{\ast}$ 
& 2.84$^{\ast}$ & 1.56 &N& 1,4\\
PG~1632$+$222  &16:34:16.09 +22:11:40.9&15.13$^{\rm b}$ &4.77 & 39384 &6.16 
& 2.84 & 1.31 &C& 1,4,8\\
HE~0001$-$2443 &00:04:31.01 $-$24:26:21.1&13.66&4.62 & 40975 & 5.94 
& 2.97 & 1.60 &N& 3\\
HZ~1  & 04:50:13.52 +17:42:06.2 &12.60$^{\rm a}$ &3.95 & 41344 & 5.68 
&3.00 & 1.87 &N& 3,4\\
            \noalign{\smallskip}
            \hline
            \hline
        \multicolumn{4}{l}{1: Also in PG survey \citep{green86}}
       & \multicolumn{6}{l}{2: Also in EC survey \citep{stob87}}\\
        \multicolumn{4}{l}{3: Also in MCT survey \citep{deme87}}
       & \multicolumn{6}{l}{4: \citet{mccook99}, misclassified as white
       dwarf}\\
       \multicolumn{4}{l}{5: \citet{chri01}}
       & \multicolumn{6}{l}{6: Possibly cool companion indicated by $B-J$,
        $J-K$ colours}\\
       \multicolumn{4}{l}{7: Possibly cool companion indicated by $B-J$ 
       colour}
       & \multicolumn{6}{l}{8: Only one spectrum available}\\
       \multicolumn{4}{l}{$^{a}$ From subdwarf data base \citep{oest06}}
       & \multicolumn{6}{l}{$^b$ from SDSS g' and r' using the 
       calibration of \citet{smi02}.}\\
       \multicolumn{4}{l}{$^c$ Photographic magnitude (SIMBAD)}
      & \multicolumn{6}{l}{$^d$ weak N lines also present}\\
\end{tabular}
\end{table*}

\subsection{Error determination \label{subsec:errors}}

\begin{figure}
\resizebox{\hsize}{!}{\includegraphics{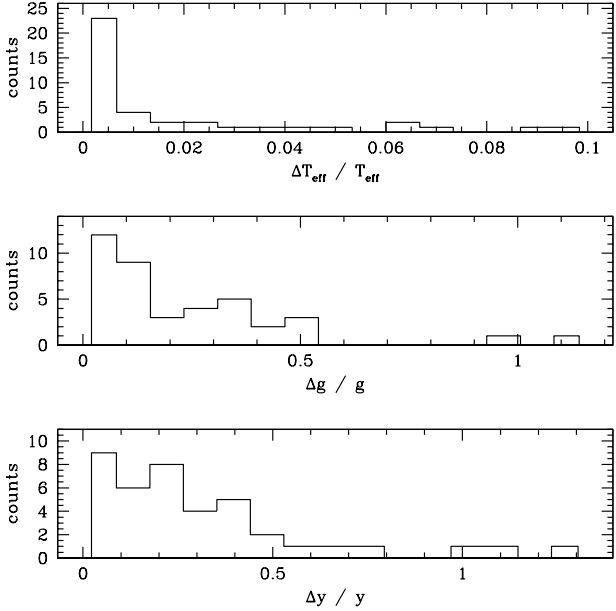}}
\caption{
Distribution of relative errors in atmospheric parameters, as derived from
two or more exposures. From top to bottom the distributions for $\teff$,
$g$ and $y=N_{He}/N_{H}$ are shown.
}
\label{error}
\end{figure}

The statistical 1-$\sigma$-errors from the fit procedure are typically
lower than 100 K, 0.04 dex and 0.04 dex for $\teff$, $\logg$ and $\logy$,
respectively, as a consequence of the high resolution and the low noise
level of our data. Nevertheless we decided to derive a more reliable
estimate of the true errors on the basis of the existing two or more
exposures of each star, where available. We determined the relative error
for  $\teff$, $g$ and $y$ by taking the difference between the fit results
of each exposure and dividing by the weighted mean of these
fit results (see Section~\ref{subsec:fit}). 
Figure~\ref{error} shows the histogram
distribution of the resulting relative errors. We find the
1-$\sigma$-value (sample standard deviation) of these distributions to
yield global relative errors of $\Delta \teff / \teff = 0.026$, $\Delta g
/ g = 0.25$ and $\Delta y / y = 0.30$. This corresponds to $\Delta
\log\teff = 0.011$, $\Delta \logg = 0.097$ and $\Delta \logy = 0.11$. We
note, that some relative errors are particularly large, i.e. more than 
2-$\sigma$ (sample standard deviation) of these distributions.
These so-called ''outliers'' are flagged in Tab.~\ref{tab_newgrid}.

\section{Spectral analysis: Results, trends and biases}\label{sec:result2}

As outlined in \citet{ede03} and paper~I we shall
search for trends in parameter space. We shall discuss the
distributions of gravity, temperature, and helium abundance, taking
into account the CN-classification scheme (see section~\ref{sec:class}).
As systematic differences in atmospheric parameters lead to selection biases, we 
discuss this issue at the end of this section.

\subsection{Atmospheric parameters and C \& N line strengths}

\begin{figure}
\resizebox{\hsize}{!}{\includegraphics{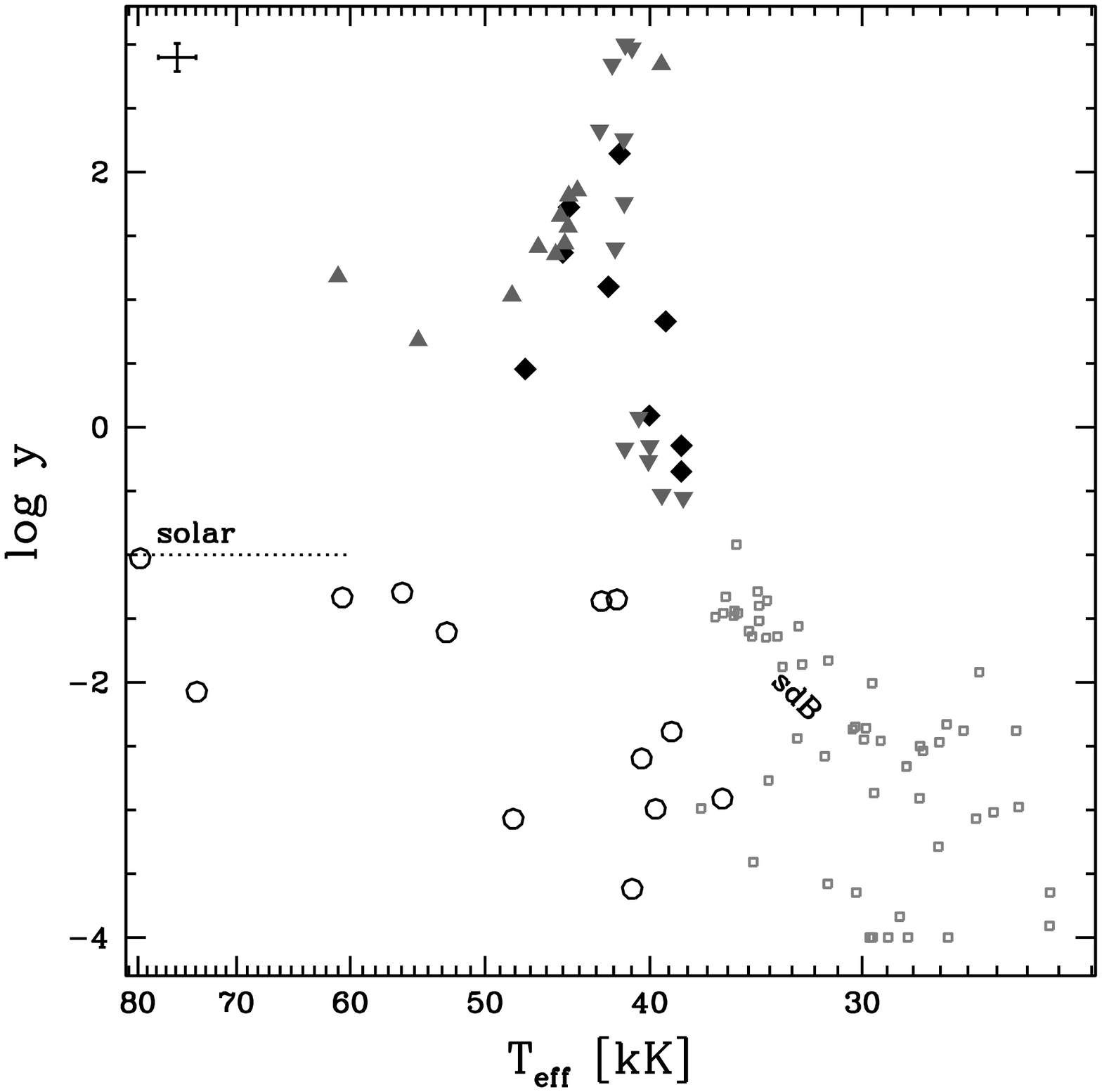}}
\caption{
Helium abundance versus effective temperature for sdO 
as shown by their CN-classification and comparison to the sdB stars (open
squares) from the 
SPY project \citep{lis05}.
Open circles denote stars
of CN class 0, filled triangles are
either class N (upside down triangle) or class C,
whereas stars of CN class are marked by a filled diamond. sdB stars are marked
by open squares.
Measurement uncertainties
are given in the upper left corner.
}

\label{teffloghe}
\end{figure}

The following conclusions can be drawn from the distribution of stars in the 
\tefflogy-diagram (Fig.~\ref{teffloghe}):

(i) SdO stars with a helium abundance below solar, ranging from $0.0003
\le He/H \le 0.08$, are scattered in a wide range 
of effective temperatures from $35\,\rm{kK} \la \teff \la
80\,\rm{kK}$, and no correlation can be found.

(ii) SdO stars with a  helium abundance exceeding the solar value,
ranging from $0.1\la He/H \la 1$,
tightly cluster around $\teff\approx 40\,\rm{kK}$ and He/H$\approx$0.5.

  (iii) Above a helium abundance of He/H=10 the sdO stars
  show a clear dependence of the helium abundance on temperature: the
   helium abundance \emph{decreases} with increasing $\teff$, 
   opposite to the general trend for sdB stars.

Most strikingly a clear correlation between helium abundance and CN class becomes apparent from 
Fig.~\ref{teffloghe}.
None of the sdO stars with subsolar helium content shows carbon and/or nitrogen lines and 
therefore all of them are of CN class 0. The opposite is true for sdO stars
with supersolar helium content -- all of them show 
carbon and/or nitrogen lines and therefore are either of class C, N, or
CN.

This suggests that the sdO stars should be grouped according to helium content
into two classes. Those with supersolar helium abundances will further on be
referred to as \emph{helium-enriched} sdO stars while those 
with subsolar helium abundances will be termed
\emph{helium-deficient}
sdO stars.

\begin{figure}
\resizebox{\hsize}{!}
{\includegraphics{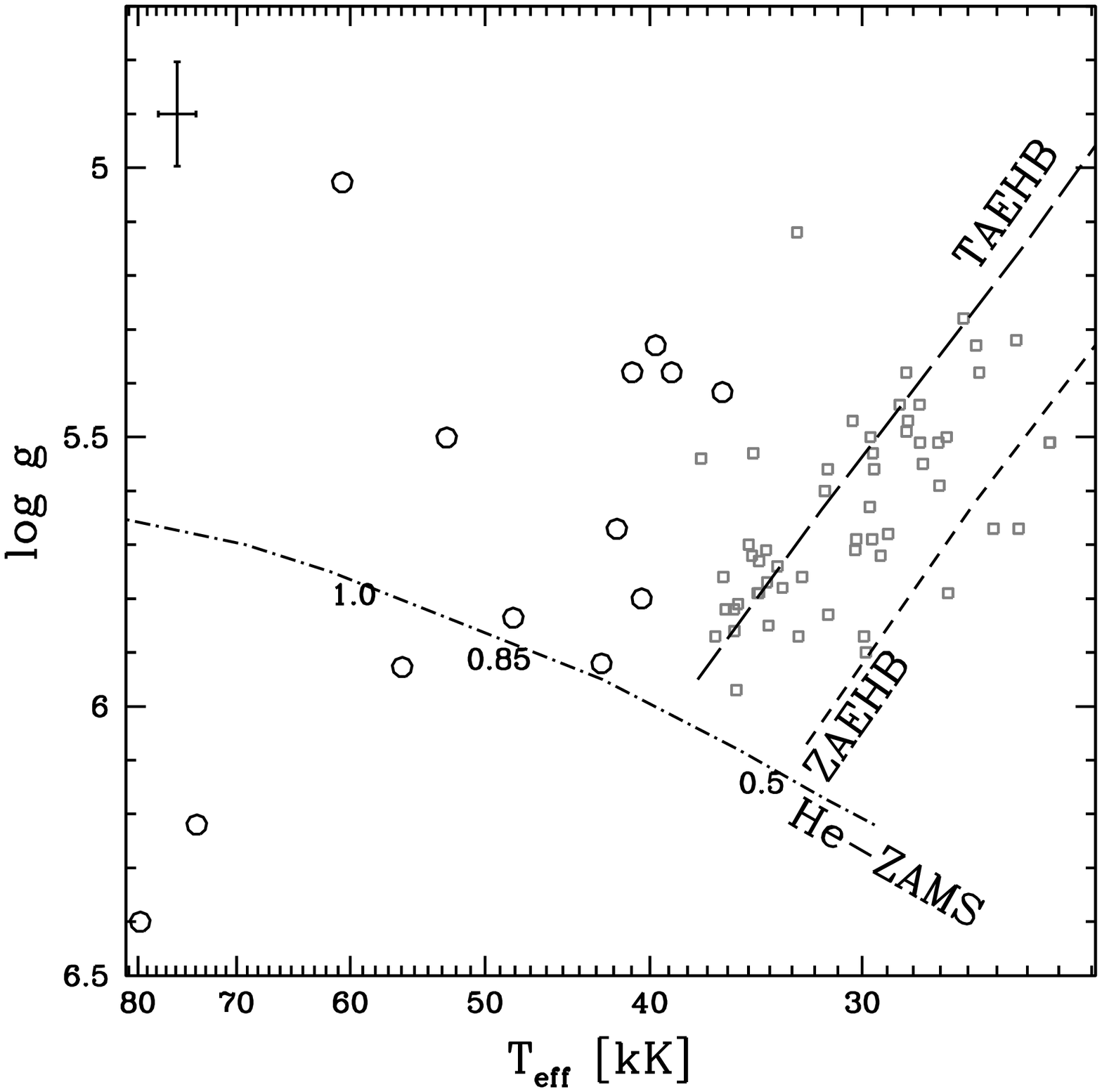}}

\caption{
{\bf Helium-deficient sdO stars}: 
Distribution of effective temperature and surface gravity.
The location of the EHB band and the helium zero-age main sequence
(He-ZAMS, labeled by stellar mass in solar units) are also 
indicated. Notation as in Fig.~\ref{teffloghe}.
Measurement uncertainties are shown in the upper left corner.
}

\label{tefflogg_sdO}
\end{figure}

\begin{figure}
\resizebox{\hsize}{!}
{\includegraphics{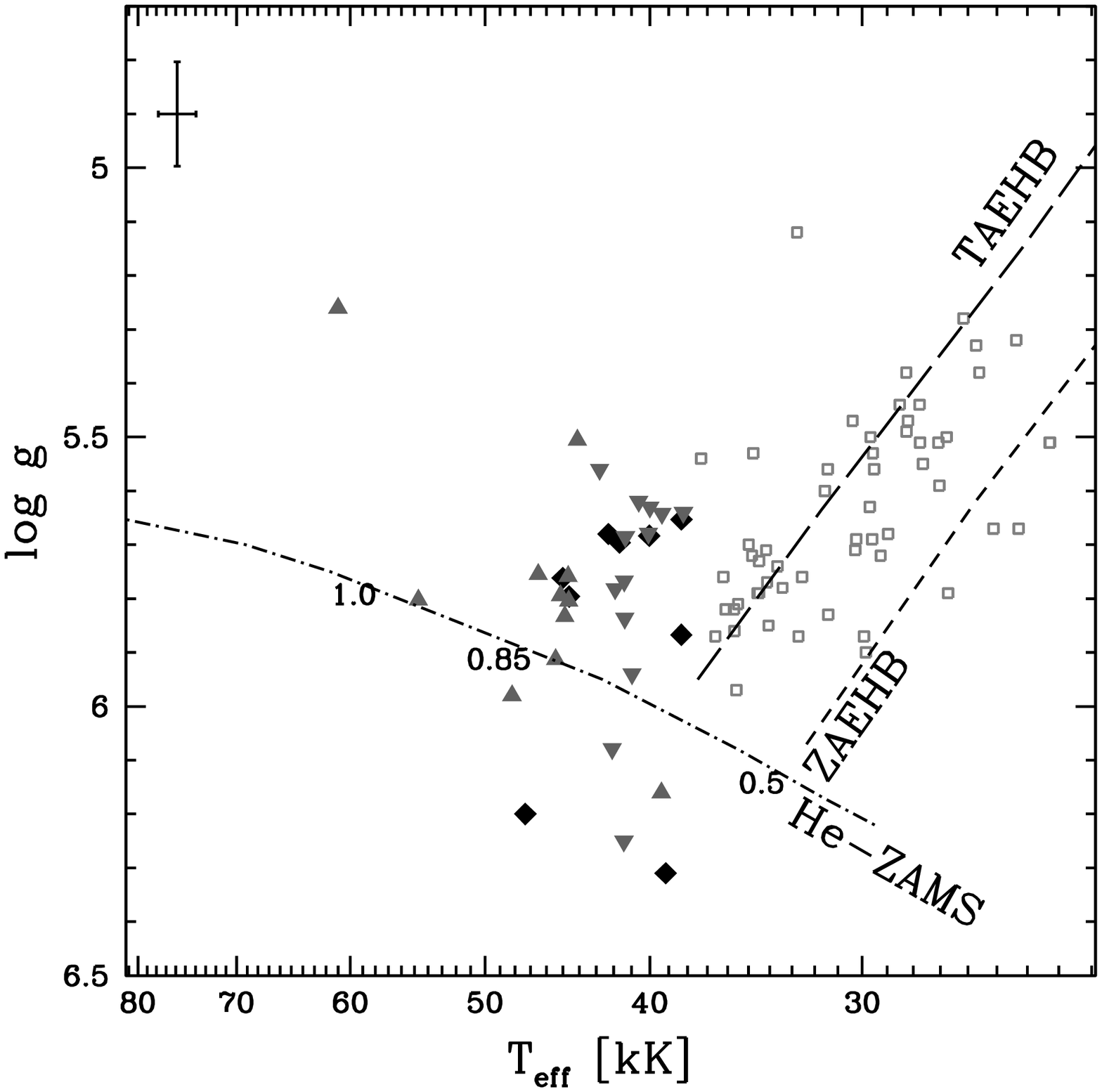}}
\caption{
{\bf Helium-enriched sdO stars}: Distribution of effective temperature and surface gravity.
The location of the EHB band and the helium zero-age main sequence
(He-ZAMS, labeled by stellar mass in solar units) are also indicated. Notation as in 
Fig.~\ref{teffloghe}. 
Measurement uncertainties are shown in the lower right corner.
}

\label{tefflogg_hesdO}
\end{figure}

Figs.~\ref{tefflogg_sdO} and ~\ref{tefflogg_hesdO} show the distribution of the full 
SPY sample of sdO stars in the \tefflogg-plane, in comparison with the sdB stars 
from paper~I.
The location of the EHB band \citep[]{dor93} is shown, along with the 
helium zero-age main sequence (He-ZAMS) which indicates the boundary
beyond which no stable helium core burning can be established
\citep[]{pac71}. 

Both \emph{helium-deficient} (Fig.~\ref{tefflogg_sdO}) and
\emph{helium-enriched}
(Fig.~\ref{tefflogg_hesdO}) sdO stars are found at higher 
temperatures than
the sdB stars, most of them between 37~kK and 47~kK.
They clearly lie outside the EHB band, which is defined as the
region between the zero-age EHB 
(ZAEHB) and the terminal age EHB (TAEHB) given by the evolutionary
calculations of \citet[]{dor93}. A significant fraction (7 out of 33) of
\emph{helium-enriched} sdOs appear to lie below the helium
main sequence. In view
of the non-Gaussian distribution of errors (see Sect.~\ref{error}), 
their gravities might have been
overestimated. Indeed, two stars are amongst the ''outliers'' and the results 
of another three are based on single spectra. 
It may therefore be premature to draw reliable conclusions. 
Additional
observations are necessary to derive more precise gravities.

While the \emph{helium-deficient} sdO stars are scattered in a wide
\tefflogg-range, most \emph{helium-enriched} sdOs populate
a relatively narrow region $(\teff$ from $\sim$40 to $\sim$46\,kK and $\logg$
from $\sim$5.5 to $\sim$5.9).

\subsection{Selection bias \label{sec:select}} 

The target objects of SPY were selected from a magnitude limited list of
candidate white dwarfs. Therefore all subdwarfs present in
the SPY dataset have only been included ``by accident'', because they were 
classified erroneously as white dwarfs mostly from low resolution
objective prism spectra. For
the sdB stars, we found in paper~I that 
any selection effects cannot be significantly different from
typical selection effects present in earlier studies of sdB stars.
Besides the usual biases of magnitude limited surveys there may be
another selection effect introduced by the SPY observing strategy.  
As the SPY project aimed at including a significant fraction of helium-rich
white dwarfs, preference was given to DB candidates, many of which turned out 
to be sdO stars.  
There may thus be a bias in favour of selecting sdO stars
\emph{relative to sdB stars} in the SPY project. 
This must be taken into account
when the 
combined sdB and sdO sample from SPY is compared to predictions from 
evolutionary calculations.  


\section{Evolutionary status \label{sec:obstheo}}

As outlined above, the distribution of the 
programme stars in the \tefflogg-diagram (see Figs.~\ref{tefflogg_sdO} 
and
 \ref{tefflogg_hesdO}) as well as with respect to CN class suggests that 
\emph{helium-enriched} sdO stars form a population different from the 
\emph{helium-deficient} sdO stars.
Published evolutionary scenarios try to explain the origin of sdB and sdO stars
either by canonical or non-canonical evolution of single stars or by close 
binary evolution
with mass exchange and common envelope episodes. We now test these scenarios by
comparing our observational results to their predictions.

\subsection{Canonical evolution of single stars}

Canonical EHB models \citep[e.g.][]{dor93} are characterized by a core mass
of nearly half a solar mass and a tiny inert hydrogen-rich envelope. The core
mass is fixed by the onset of the core helium flash at the tip of the red giant
branch and depends only slightly on metalicity and helium abundance. Hence the 
canonical core mass is restricted to a very narrow range of 0.46 to 0.5~$\msol$.
This configuration prevents an EHB star from ascending the AGB. The post-EHB
evolution proceeds towards higher temperatures until the white dwarf cooling 
track is reached and gravity increases. 

The problem for the formation of EHB
stars in this case is, how almost the entire envelope of the RGB progenitor
is lost at precisely the same time as the core reaches the mass required for the
helium flash. Enhanced mass-loss during or after the red giant branch has been
postulated \citep{dcr96} but no physical mechanism has yet been identified. 

\begin{figure}
\centering
\resizebox{\hsize}{!}{\includegraphics{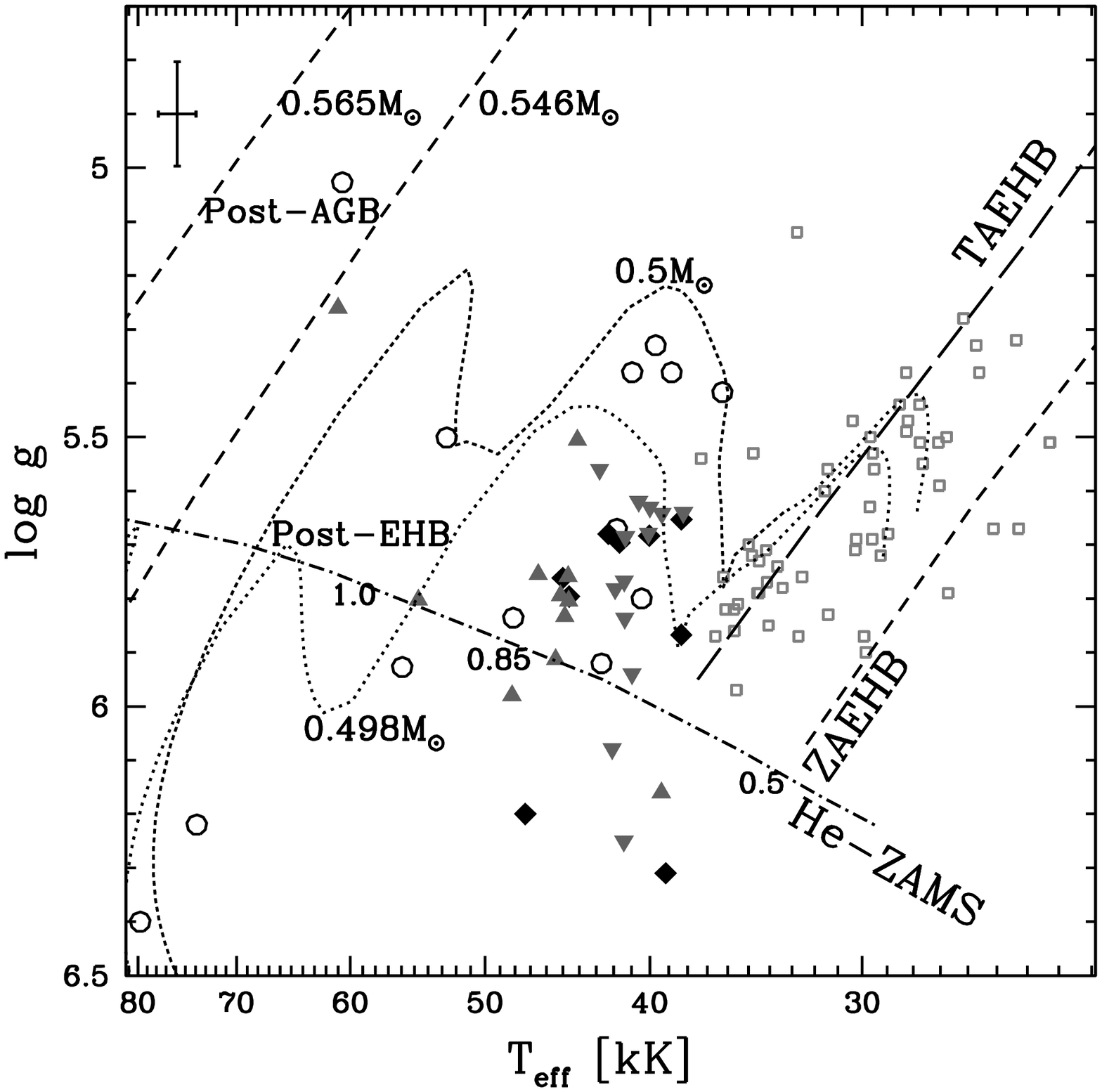}}

\caption{
Tracks for post-EHB evolution (black dotted lines) from
\citet[]{dor93} are superposed onto the \tefflogg-diagram for \emph{helium-deficient} 
and \emph{helium-enriched} sdO stars, with core
masses as indicated next to the tracks. 
Notation as in Figs.~\ref{tefflogg_sdO} and \ref{tefflogg_hesdO}. 
We also show post-AGB tracks
(dashed lines) from \citet[]{schoe79,schoe83} with white dwarf
(final stage) masses as indicated next to the tracks.
The helium main sequence is labeled by stellar mass in solar units.}
\label{cI}
\end{figure}

In Fig.~\ref{cI} we compare the distribution of sdO stars to the position of the
EHB-band and to post-EHB
evolutionary tracks and find that none of the programme stars lies on the 
EHB-band. However, it is premature to conclude that they have evolved 
from the EHB. Most of the \emph{helium-enriched} sdO stars cluster in a narrow region
of the \tefflogg-diagram. The calculations of \citet{dor93} indicate that the 
pace of evolution does not change very much through post-EHB evolution. 
Hence post-EHB evolution can not explain the clustering of 
\emph{helium-enriched} sdO
stars. Moreover, the post-EHB scenario fails to explain the surface enrichment 
of helium and it also can not explain why the C and/or N lines 
in \emph{helium-enriched} sdO stars are stronger
than \emph{helium-deficient} ones, since no dredge-up process is predicted to occur.

The \emph{helium-deficient} sdO stars, however, can well be explained by the 
post-EHB hypothesis
as they have similarly low helium abundances as sdB stars. In addition they 
are scattered in the \tefflogg-diagram, hence no slow-down in their evolution 
has to be invoked.
Therefore it is reasonable that \emph{helium-deficient} sdO stars are  
post-EHB stars and have evolved from sdB stars, 
while \emph{helium-enriched} sdO stars have not.

Some of the more luminous sdO stars may not be related to the EHB at all but 
may have evolved off the AGB. Therefore we included 
post-AGB evolutionary tracks from \citet{schoe79,schoe83} in Fig.~\ref{cI}. 
The position of the \emph{helium-deficient}
sdO HE~1047$-$0637 ($\teff\approx 60\,000K, \logg\approx 5.0$) 
is matched 
by these tracks,
indicating a possible post-AGB evolutionary stage
while the \emph{helium-enriched} sdO HE~1430$-$0815 
($\teff\approx 61\,000K, \logg\approx 5.3$) may be either a post-AGB or
a post-EHB star. The short
evolutionary timescales \citep[about 30\,000 years from the AGB towards a
pre-white dwarf,][]{schoe79,schoe83} drastically reduce the probability  
of finding true post-AGB stars.

\subsection{Non-canonical evolution of single stars}


Since canonical single star evolution can not explain the \emph{helium-enriched} sdO
stars, in particular their clustering in the \tefflogg-plane, the
helium enrichment and the C and/or N line strengths, other scenarios have
to be investigated, such as
the late hot flasher scenario, in which the core helium flash occurs when the 
star has already left the RGB and is approaching the white 
dwarf cooling sequence (delayed helium core flash). During the flash, He 
and C is dredged-up to the surface \citep[]{swe97}.
\begin{figure}
\centering
\resizebox{\hsize}{!}{\includegraphics{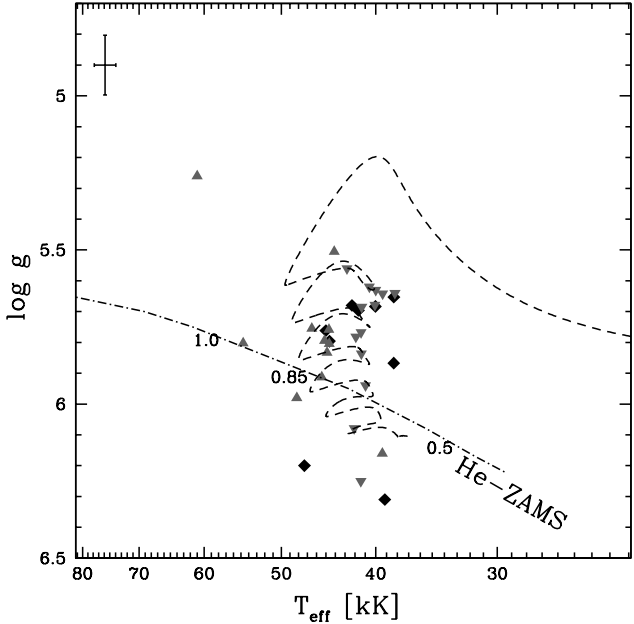}}
\caption{
{\bf Helium-enriched sdO stars}: comparison with an evolutionary track for an 
EHB star formed by a delayed helium flash in the 
effective temperature vs. surface gravity plane. The track settles onto
the helium main sequence (labeled by stellar mass in solar units). 
Symbols as in Fig.~\ref{tefflogg_hesdO}.
}
\label{latehe}
\end{figure}

Fig.~\ref{latehe} compares an evolutionary track
 for the late hot flasher scenario with
the distribution of our {\emph helium-enriched} sdO stars.
The star encountered a late flash on its way  
from the RGB towards the white dwarf regime. This late hot flash forces the 
star to land on or near the helium main sequence, i.e. at the extreme end of the
EHB. 
The final composition of the envelope is helium dominated with Y=0.814, 
complemented by hydrogen (X=0.154) and enriched with carbon at Z=0.032 
\citep[or nitrogen if the hydrogen burning during the helium
flash phase burns $^{12}$C into $^{14}$N; ][]{swe97}. 
Indeed, most of our observed \emph{helium-enriched} sdO 
stars lie near the model track, suggesting that 
they indeed may  
originate from this scenario \citep[see also][]{lemk97}. 
However, the evolutionary time scales 
(1.95$\times 10^6$ yrs for the evolution shown in Fig.~\ref{latehe}; 
\citealt{swe97}) are much shorter than for the core helium burning phase. 
Accordingly the stars should accumulate near the end of the track, i.e. near the
helium main sequence, which is not the case for our programme stars. 

Although the late hot flasher scenario can explain the helium enrichment and 
the line strengths of C and/or N lines as due to dredge up, it fails to
reproduce the distribution of the stars in the \tefflogg-diagram in detail.
 
\subsection{Binary evolution \label{sec:evol_hesdo}}

As the evolutionary scenarios for single stars discussed above partially 
fail to
explain
the observed properties of \emph{helium-enriched} sdO stars, we now focus on
binary evolution. 
Two flavours of close binary evolution have been envisaged to explain sdB
and sdO
stars. The formation of core helium burning EHB stars through close binary 
evolution has extensively been investigated by \citet{han02,han03}
using the
binary population synthesis approach.

However, it may not be taken for granted that hot subluminous stars are 
indeed core helium burning. They may form from red giants that left the RGB
before igniting helium in the core and evolve through the EHB region 
as helium stars towards the white dwarf cooling sequence, see \citet{heb03bb}. 
Such progenitors of helium core white dwarfs, indeed, have been discovered. 
\citet{heb03a} found that the sdB star HD~188112 has a mass of 0.23$\msol$, 
too small to sustain helium burning. 

\subsubsection{Binary evolution scenarios and the role of white dwarf
mergers}

\citet[][hereafter HPMM]{han03} 
showed in their binary population synthesis study that three
 channels are relevant for the formation of hot subluminous stars involving 
 either 
common-envelope ejection, stable Roche lobe overflow or a merger of two
helium white dwarfs. Stable Roche lobe overflow is predicted to lead to 
composite spectrum systems consisting of a hot subdwarf and a main sequence
star. 
There are eight such binaries (sdO + main sequence star) present amongst
our programme stars. 

In paper~I, we compared the atmospheric parameters of the
SPY sdB stars to the HPMM models by using two diagnostic tools, namely the
\tefflogg-diagram and the cumulative luminosity function.
Our analysis of sdO stars extends the subdwarf
sample to higher temperatures which may allow to study the link between 
sdB and sdO stars. 
One would thus expect a comparison of our full sdB/sdO
sample with the HPMM models to yield more robust results than in paper~I.
However, we must observe selection biases as discussed in 
section~\ref{sec:select}.

\begin{figure}
\centering
\resizebox{\hsize}{!}{\includegraphics{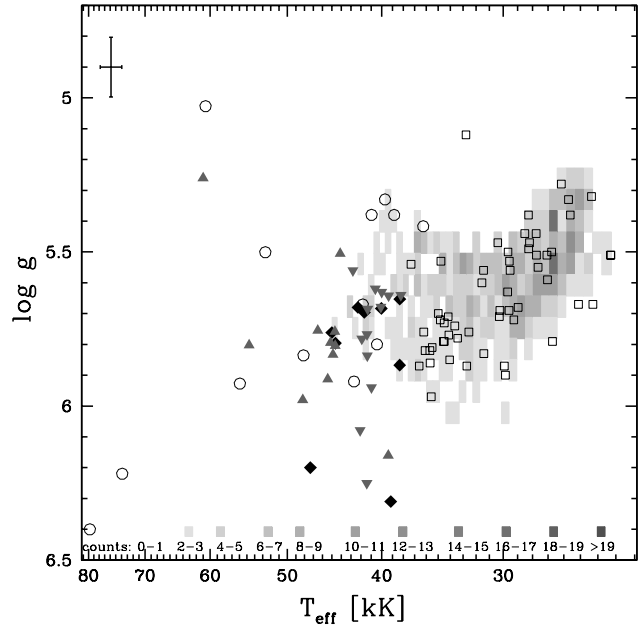}}
\caption{
Comparison of the atmospheric parameters of \emph{helium-deficient} sdO,
\emph{helium-enriched}
sdO, and sdB
from the SPY project
to simulation set No. 10 of \citep{han03}, which was found to match the sdB
distribution of Paper~I best. The
theoretical predictions are shown as shaded
\tefflogg-boxes, where a higher subdwarf density per box corresponds to
darker shading. The grey scale is shown below the
figures. Notation as in Fig.~\ref{teffloghe}.
}

\label{hpmm1}
\end{figure}

We present this comparison for the
\tefflogg-diagram in Fig.~\ref{hpmm1}. The simulation set No. 10 of HPMM was
chosen because it came closest to the SPY-sdB distribution
(see paper~I).  
The grey shading of the rectangular areas corresponds to the respective
number of simulated stars they contain (cf.\ paper~I). Higher
number densities of simulated subdwarfs correspond to darker grey
shading. 
We refer the reader to paper~I for more details about the general 
specifications of the HPMM simulations.

From the direct comparison of our derived \tefflogg-values to the HPMM
simulations, we see two effects. First, sdO stars significantly exceed
even the hottest
temperatures that result from any HPMM simulation set, with stars
reaching up to 80 kK. Even for the sdO stars with
lower temperatures, no set is able to reproduce their rather wide
range in surface gravity. HPMM therefore covers only $\le 38$\% 
 of all sdO stars. 
Second, by restricting our 
analysis to stars which come close to the HPMM predictions, i.e.
those that are apparently connected with the sdB sample, a strong
disagreement of the observational data with the simulation set becomes
obvious: the relative amount of hot (sdO) and cool (sdB) stars differs
significantly. Apart from possible limitations of the HPMM models (as
discussed below in section \ref{sec:discuss}), there may be an
observational bias in favour of sdO stars, rendering
a \emph{quantitative} 
comparison with HPMM, as performed for the sdB stars in paper~I, very 
difficult or even impossible.

Let us now focus on the white dwarf merger channel, which is supposed to 
result in stars with very low hydrogen envelope mass. Hence stars resulting from mergers of two helium-core white dwarfs are 
expected to be found near the hot end of the HPMM distribution.
Mergers can also 
produce stars in a wider range of masses than the other channels predict. 
The merging process probably induces lots of mixing of nuclear processed 
material to the stellar surface of the remnant, potentially leading to a
helium- and nitrogen-rich surface composition. Depending on the efficiency of 
nuclear burning and mixing, carbon may also be enriched at the surface.
As these predictions match the observed properties of the \emph{helium-enriched} 
sdO stars at least qualitatively, we regard the merger of helium core white
dwarfs as a viable scenario. 
There are two predictions we can test: (i) Stars formed from a merger should not
be radial velocity variable (unless they stem from triple systems) and 
(ii) the helium enrichment is accompanied by enrichment of nitrogen and/or
carbon. While our CN-classification scheme (section~\ref{sec:class}) is based on the 
\emph{presence} of C and/or N-lines in the spectrum, the latter issue requires 
a \emph{quantitative} abundance analysis. 
We shall address these tests in forthcoming papers.     
 
\subsubsection{Non-core helium burning stars in close binary systems} 

The evolution of RGB
stars whose envelopes
get almost completely stripped by Roche lobe overflow in a close
binary system \emph{before} helium burning starts has been investigated 
e.g. by 
\citet{dri98}. These tracks were calculated from a 1$\msol$
model sequence starting from the pre-main sequence stage up through the
RGB. Large mass loss rates were then adopted and the evolution of the resulting
helium star was followed.
The remnant finally evolves into a helium core white dwarf.

\begin{figure}
\centering
\resizebox{\hsize}{!}{\includegraphics{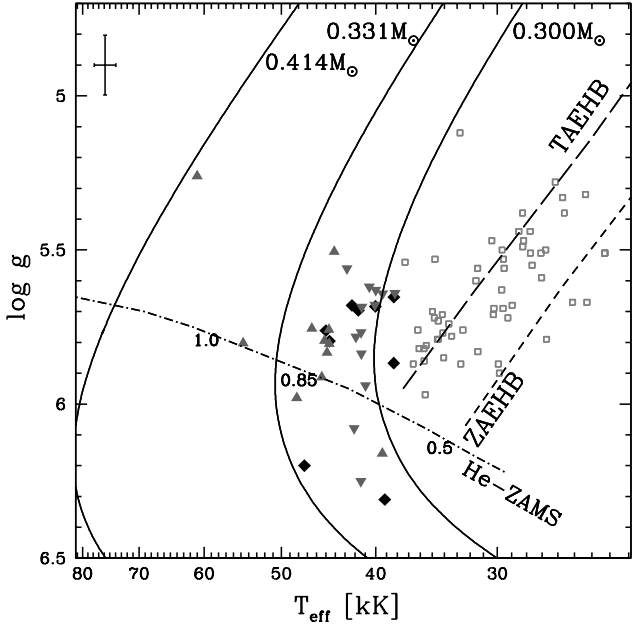}}
\caption{
Tracks for post-RGB evolution (solid lines) towards helium core white dwarfs 
of 0.3$M_{\odot}$ to
0.414M$_{\odot}$ \citep[]{dri98} are superposed onto the
\tefflogg-diagram of \emph{helium-enriched} sdO stars,
with masses of the
model star as indicated next to the tracks. 
The helium main sequence is labeled by stellar mass in solar units.
Notation as in Fig. \ref{teffloghe}.
}
\label{driebe}
\end{figure}

In Fig.~\ref{driebe} we compare the programme stars to the predictions of 
the evolutionary models of \citet{dri98} for different masses. 
While the position in the \tefflogg-diagram of any of our programme stars
can be matched by a post-RGB track of appropriate mass, it is striking that 
most of the \emph{helium-enriched} sdO stars agree reasonably well with the 
theoretical predictions for a rather narrow mass range, i.e.
 between 0.3M$_{\odot}$ and
0.33M$_{\odot}$. Hence the distribution of \emph{helium-enriched} sdO stars 
could be a sequence of low mass stars evolving into helium white dwarfs. 
It is also surprising that the predicted masses are close to the minimum
mass for the helium main sequence, which might be purely coincidental. 

However, the models of \citet{dri98} predict rather thick hydrogen layers.
Therefore, this scenario has difficulties to explain the high helium 
abundances at the stellar surface as observed in \emph{helium-enriched} sdO stars.
During the post-RGB evolution hydrogen-shell flashes occur which possibly may 
lead to a dredge up of helium and nitrogen \citep{dri99}. 
This may explain nitrogen strong-lined objects. 
It is, however, not evident how carbon-strong objects
could be formed. Moreover, the models of \citet{dri99} indicate that the
occurrence of hydrogen flashes is constrained to a lower mass range of 
0.21 to 0.3$\msol$. 

If the hydrogen were removed from the envelope by some unknown process, the 
evolution would be speeded up considerably, 
because most of the luminosity is provided by the
hydrogen burning shell, which would be extinguished if the hydrogen mass is too
low. Rapid evolution reduces the detectability of such stars. 
 
Therefore we regard the post-RGB scenario as unlikely. 
However it is a viable scenario to explain some of the helium-\emph{deficient} 
sdO stars since
helium core white dwarf progenitors are known albeit rare amongst the sdB stars 
\citep[e.g. HD~188112,][]{heb03a}.      

\section{Summary and Discussion \label{sec:discuss}}

We analyzed  high resolution optical spectra of 58 subluminous O stars.
We found spectroscopic and/or photometric evidence 
for cool companions to eight subluminous O stars.
While HE~1502$-$1019 and HE~1513$-$0432 display only weak helium lines, 
the others have strong helium lines. 
HE~0301$-$3039 is an unique binary consisting of two sdO stars 
with spectra dominated by helium lines
\citep{lis04}. Evidence for cool companions to the other helium-enriched 
stars rests on 
photometry only. Near-infrared spectroscopy is required for classifying them. 
As the photometric measurements for our sample are only available for 52 out of
58 stars, the 
fraction of sdO stars with cool companions may be slightly larger.

A grid of synthetic H-/He-line spectra calculated from
NLTE model atmospheres was used to derive the stars'
atmospheric parameters. 
Reliable atmospheric parameters were obtained for 13 \emph{helium-deficient} sdO
and 33 \emph{helium-enriched} sdO stars. 
A clear-cut correlation between CN class and helium abundance was found.
\emph{Helium-deficient} sdO stars did not show any C and/or N lines, while 
all \emph{helium-enriched} sdO stars do show either C or N lines or both, 
suggesting
that they form two different populations of stars.
The \emph{helium-deficient} sdO stars are scattered
in the \tefflogg\ diagram, whereas most \emph{helium-enriched} sdO stars cluster in a 
narrow range.

Comparing the observed distribution to the predictions of evolutionary
calculations for single as well as for close binary stars, 
we conclude that many  \emph{helium-deficient} sdO stars can be explained as evolved 
sdB stars.
Both classes of star are helium deficient and have weak metal lines which is
caused by atmospheric diffusion processes.

Most of the \emph{helium-enriched} sdO stars cluster in a narrow region
of the \tefflogg-diagram at temperatures between 40kK and 50kK.
While diffusion is probably causing helium deficiency,
it is unlikely to account for the helium enrichment. Non-standard
evolutionary scenarios were therefore considered as well.
The predictions from the late hot flasher scenario as well
as the helium white dwarf merger scenario are roughly consistent with
the observed distribution of \emph{helium-enriched} sdO stars but do not
match them in detail. 
The occurrence of a delayed helium core flash as well as the merger of two
helium white
dwarfs may explain the helium enrichment.
In both cases carbon and/or nitrogen
can be dredged up to the stellar surface, which would
explain the strength of the C and/or N lines in \emph{helium-enriched} sdO 
stars.

Some high gravity \emph{helium-enriched} sdO stars may lie
below the helium main sequence, which is at variance with any core helium 
burning
model. Therefore, we considered models for post-RGB
stars with inert helium cores which evolve through the sdB/sdO regime in the
\tefflogg\ diagram into helium core white dwarfs. This is the only scenario
that can explain stars to lie below the helium main sequence. However, it is
not obvious how the helium enrichment is brought about as the hydrogen
envelopes of post-RGB models has to be relatively thick or, otherwise, 
the stars would evolve too fast to be observable in large quantities.
If the existence of a population of high gravity sdO stars below the 
helium
main sequence could be confirmed by surveys with larger sample size, a
possible explanation could be provided by the post-RBG scenario.

Our conclusions can be tested by measuring the binary frequency and the C and
N abundances.
If \emph{helium-deficient} sdO stars are evolved sdB stars, their binary
fractions should be the same. As about 40\% of the sdB stars are in close
binaries with periods below 10 days \citep{napi04},
we expect 5 \emph{helium-deficient} sdO stars in our sample to be radial
velocity variable on time scales of 10 days or less. An investigation of the binary frequency 
is underway in our group.
If \emph{helium-enriched} sdO stars result from mergers, no radial velocity variations
would be expected (except for objects arising from triple stars).
If material processed by nuclear burning is dredged up, the helium enrichment
is expected to be accompanied by enrichment of nitrogen and/or carbon.
While our CN-classification scheme (section~\ref{sec:class}) is based on the 
\emph{presence} of C and/or N-lines in the spectrum, the latter issue requires 
a \emph{quantitative} abundance analysis to test the
late hot flasher and the merger scenario. 

In addition, improvements to our analyses of the SPY-sample are required
for a more detailed comparison with predictions of evolutionary calculations.
The role of metal-line blanketing in the NLTE model atmospheres needs to be
investigated. Due to the complexity of the problem, only few studies of
NLTE metal line blanketing in sdO star atmospheres
are available \citep{haas96,lanz97,deet00} which
indicate that the effective temperatures may have to be reduced. According
to these investigations, 
the most important contribution to line blanketing stems from the iron group
elements. However, their abundances can be derived from ultraviolet
spectra only.
The observed diversity of carbon and nitrogen line strengths indicates
that the treatment of metals has to be done carefully on a 
star-by-star basis.

Apparently, the SPY sample suffers from observational selection biases.
We found it likely that sdO stars were selected preferentially
for the SPY target list. Hence sdO stars may be overrepresented with 
respect to sdB
stars.
To this end the Sloan 
Digital Sky Survey (SDSS) is a promising source of hot
subdwarf stars, as its selection criteria are very different from that of
SPY.
We have already begun a spectroscopic analysis of SDSS sdO stars
\citep{heb06} which 
will contribute significantly towards discriminating between the
various evolutionary scenarios and hypotheses outlined above.

\begin{acknowledgements}
T.L. gratefully acknowledges support by the Swiss National Science
Foundation. R.N. is supported by a PPARC Advanced Fellowship.
We thank Iris Traulsen and Thomas Rauch for their help in running the 
model atmosphere codes, Heiko Hirsch for double checking the
CN-classification, and Roy {\O}stensen for creating 
the subdwarf data base \citep{oest06}, which we used extensively.
This publication makes use of data products from the Two Micron All Sky
Survey, which is a joint project of the University of Massachusetts and
the Infrared Processing and Analysis Center/California Institute of
Technology, funded by the National Aeronautics and Space Administration
and the National Science Foundation \citep{2MASS}.
This publication makes use of the VizieR database of astronomical
catalogs \citep{Vizier}.
\end{acknowledgements}

  \bibliography{5564}
  \bibliographystyle{aa}


\end{document}